\documentclass[reqno,12pt]{amsart}

\usepackage{amsmath,amssymb,amsthm,amsfonts,amsxtra}
\usepackage{fullpage}

\begin{document}

\newtheorem{thm}{Theorem}[section]
\newtheorem{cor}{Corollary}[section]
\newtheorem{lem}{Lemma}[section]
\newtheorem{prop}{Proposition}[section]
\theoremstyle{definition}
\newtheorem{exmp}{Example}

\def\Ref#1{Ref.~\cite{#1}}

\def\const{\text{const.}}
\def\Rnum{{\mathbb R}}
\def\sgn{{\rm sgn}}
\def\Dop{{\mathcal{D}}}
\def\nvec{{\mathbf n}}
\def\kvec{{\mathbf k}}
\def\sech{{\rm sech}}

\def\sig{\sigma^2}
\def\sqrtsig{\sigma}
\def\const{\text{const.}}
\def\t{{\rm t}}

\def\Esp{{\mathcal E}}
\def\X{{\rm X}}
\def\pr{{\rm pr}}

\def\z{\zeta}
\def\barT{\bar T}
\def\barPhi{\bar\Phi}
\def\barTheta{\bar\Theta}

\def\barG{\bar G}
\def\barQ{\bar Q}
\def\x{\mathbf{x}}
\def\A{\mathbf{A}}

\tolerance=10000
\allowdisplaybreaks[3]

\title{Symmetry multi-reduction method for partial differential equations with conservation laws}

\author{
Stephen C. Anco$^1$
\lowercase{\scshape{and}}
M.L. Gandarias${}^2$
\lowercase{\scshape{}}
\\\\\lowercase{\scshape{
${}^1$Department of Mathematics and Statistics\\
Brock University\\
St. Catharines, ON L2S3A1, Canada}} \\\\
\lowercase{\scshape{
${}^2$Department of Mathematics\\
Faculty of Sciences, University of C\'adiz\\
Puerto Real, C\'adiz, Spain, 11510}}\\
}


\thanks{${}^1$sanco@brocku.ca, ${}^2$marialuz.gandarias@uca.es}

\begin{abstract}
For partial differential equations (PDEs) that have $n\geq2$ independent variables
and a symmetry algebra of dimension at least $n-1$,
an explicit algorithmic method is presented for
finding all symmetry-invariant conservation laws that will reduce to first integrals for the ordinary differential equation (ODE)
describing symmetry-invariant solutions of the PDE. 
This significantly generalizes the double reduction method known in the literature. 
Moreover, the condition of symmetry-invariance of a conservation law
is formulated in an improved way by using multipliers, 
thereby allowing symmetry-invariant conservation laws to be obtained directly, 
without the need to first find conservation laws and then check their invariance.
This cuts down considerably the number and complexity of computational steps involved in the reduction method. 
If the space of symmetry-invariant conservation laws has dimension $m\geq 1$,
then the method yields $m$ first integrals
along with a check of which ones are non-trivial via their multipliers.
Several examples of interesting symmetry reductions are considered:
travelling waves and similarity solutions in $1+1$ dimensions;
line travelling waves, line similarity solutions, and similarity travelling waves in $2+1$ dimensions;
rotationally symmetric similarity solutions in $n+1$ dimensions. 
In addition, examples of nonlinear PDEs for which the method yields 
the explicit general solution for symmetry-invariant solutions are shown. 
\end{abstract}

\maketitle

\section{Introduction}\label{sec:intro}

A powerful application of symmetries is finding group-invariant solutions of nonlinear partial differential equations (PDEs). 
These solutions satisfy a reduced differential equation (DE) in fewer variables
given by the invariants of the chosen symmetry group.
Nevertheless, this reduced DE must be solved to obtain the group-invariant solutions in an explicit form,
which requires finding sufficiently many first integrals to reduce its order so that a quadrature is obtained. 
If the reduced DE is an ordinary differential equation (ODE),
then a further reduction of order occurs 
whenever the starting nonlinear PDE has a Lagrangian formulation
in which the symmetry commutes with a variational symmetry.
Noether's theorem then provides a local conservation law
which reduces to a first integral of the ODE. 
This gives a double reduction of order for the variational PDE
and can be viewed as a counterpart of the well known
double reduction of variational problems for ODEs \cite{Whi-book,Olv-book}.

A more general double-reduction method \cite{Sjo2007,Sjo2009}
which applies to non-variational PDEs has been developed a decade ago.
It involves finding a symmetry that leaves invariant the conserved current in a local conservation law of a given PDE.
If the reduction of the PDE under this symmetry is an ODE,
then the corresponding reduction of the conserved current yields a first integral of the ODE.
When the given PDE arises from a Lagrangian,
the resulting first integral is the same as the one obtained by
the Lagrangian reduction method. 

The double reduction method has a straightforward extension \cite{BokDweZamKarMah}
to PDEs in any number of independent and dependent variables,
where the reduced DE is a PDE with one less independent variable
and the reduction of an invariant conserved current yields a conservation law of the reduced PDE.
There is a further extension of the method to considering conserved currents that are more generally invariant modulo a trivial current \cite{Sjo2009}.
This corresponds to requiring that the underlying local conservation law,
rather than the current current itself, is invariant \cite{Anc2016,AncKar}. 

Many applications of double reduction to physically interesting PDEs
have been studied
\cite{Sjo2007,Sjo2009,doublereduc1,doublereduc2,doublereduc3,doublereduc4,doublereduc5,doublereduc6,doublereduc7,doublereduc8,doublereduc9,doublereduc10,doublereduc11,AncPrz}. 
In all of these applications,
the starting point is a PDE with a known conservation law and a known symmetry group.
The method consists of finding a particular symmetry generator
under which the conserved current of the conservation law either is strictly invariant 
(sometimes called the ``symmetry associated to the conservation law'' \cite{KarMah2000,KarMah2002})
or is invariant modulo a trivial current
(properly called symmetry-invariance of the conservation law \cite{Anc2016,AncKar}).
However, unless the given PDE is of second order,
then double reduction does not provide enough first integrals to obtain a quadrature of the ODE for symmetry-invariant solutions of the PDE. 

The purpose of our work is to present a different, improved, more general approach to symmetry reduction of PDEs with conservation laws. 
\begin{itemize}
\item
First,
rather than start from a conservation law and seek symmetries under which it is invariant,
we instead start from a symmetry to be used for reduction of a PDE, 
and then find all conservation laws that are invariant under the symmetry;
each one will be inherited by the reduced DE.
This extension is particularly useful when a PDE in two independent variables is being reduced to an ODE,
as then a set of first integrals can be obtained
which allows for further reduction of the ODE.
\end{itemize}
\begin{itemize}
\item 
Second,
we consider reduction of a PDE under an algebra of symmetries rather than a single symmetry. 
For a sufficiently large symmetry algebra,
this generalization allows for direct reduction of a PDE in three or more independent variables to an ODE plus a set of first integrals.
\end{itemize}
\begin{itemize}
\item 
Third,
we formulate the condition of symmetry-invariance of a conservation law
in an improved way by using multipliers, 
which allows symmetry-invariant conservation laws to be obtained directly and algorithmically, without the need to first find conservation laws and then check their invariance.
This cuts down considerably the number and complexity of computational steps involved in the reduction method. 
\end{itemize}

We illustrate these new results by looking at a variety of physically interesting examples of
symmetry reduction of a PDE to an ODE plus a set of first integrals.
Since each first integral yields a further reduction of the ODE,
we directly obtain a \emph{multi-reduction} of the PDE under the symmetry,
without having to do successive single reductions. 

In section~\ref{sec:1+1}, we consider reduction of
a compacton equation under translation (travelling waves),
and a damped Boussinesq equation under scaling (similarity solutions). 
These first examples show
how reduction of PDEs in $1+1$ dimensions under a single symmetry
can yield more than one first integral,
and how the reduction is carried out in an improved way via multipliers. 

In section~\ref{sec:2+1}, we consider
reduction of a generalized KP (Kadomtsev-Petviashvili) equation under two translations (line solitons), 
an anisotropic thin film equation and a semilinear wave equation
under a translation and a scaling
(line similarity and travelling similarity solutions), 
in $2+1$ dimensions. 
These examples show how reduction works for a solvable algebra of symmetries,
and how the reduction is carried out all at once without having to do successive single reductions. 

In section~\ref{sec:n+1}, we consider in $n+1$ dimensions
the reduction of a porous media equation under rotations and scaling. 
This last example shows how reduction is possible under a non-solvable algebra of symmetries.

A brief outline of the multi-reduction theory needed for the examples
is given at the beginning of each section.
The general theory with full details and proofs
will be presented elsewhere \cite{AncGan2019}. 
For a general review of symmetries, conservation laws and multipliers,
from the viewpoint used in our work,
see Refs.~\cite{Anc-review,Anc2016,AncKar,Olv-book,BCA-book}.
Related background can be found in Refs. \cite{AncBlu2002b,Wol,Anc2003}. 

Finally, we make a few concluding remarks in section~\ref{sec:remarks}.

\section{Multi-reduction under a single symmetry}\label{sec:1+1}

For PDEs in $1+1$ dimensions,
reduction under a single symmetry yields an ODE.
We will show how to obtain the symmetry-invariant conservation laws of the PDE and their direct reduction to first integrals of the ODE.
This leads to a further reduction of the ODE by the number of first integrals that are functionally independent.

We will explain the method first for a translation symmetry
by considering reduction to a travelling wave ODE,
since this is perhaps one of the most common applications of symmetry reduction.
It is well known that all other infinitesimal point symmetries
can be expressed in a canonical form as a translation symmetry,
after a point transformation on the independent and dependent variables. 
Nevertheless,
since another very common application of symmetry reduction involves
similarity ODEs obtained by scaling reduction,
we will explain our method separately for this case too,
without having to make a point transformation to a translation symmetry.
This will also provide a guide as to how our method works in general 
for other point symmetries. 

To begin, we review some essential general aspects of conservation laws, multipliers, and symmetries in $1+1$ dimensions.  

Consider a scalar PDE $G(t,x,u,u_t,u_x,\ldots)=0$ for $u(t,x)$. 
A local conservation law is a continuity equation 
$D_t T + D_x \Phi =0$
holding on the space, $\Esp$, of solutions of the PDE,
where $T$ is the conserved density and $\Phi$ is the spatial flux,
which are functions of $t,x,u$, and derivatives of $u$.
Here $D_t,D_x$ are total derivatives, and $(T,\Phi)$ is the conserved current.

A conserved current is trivial if, on the solution space $\Esp$, 
it has the form $T|_\Esp=D_x\Theta$ and $\Phi|_\Esp=-D_t\Theta$,
with $\Theta$ being a function of $t,x,u$, and derivatives of $u$.
Trivial conservation laws contain no useful information about the PDE or its solutions,
and hence only non-trivial conservation laws are of interest. 

Every non-trivial conservation law, up to addition of a trivial conservation law,
arises from a multiplier $Q$ which is a function of $t,x,u$, and derivatives of $u$
such that $Q|_\Esp$ is non-singular
and $QG =D_t T + D_x\Phi$ holds as an identity. 
Under a mild regularity condition \cite{Olv-book,Anc-review}
on the form of the PDE $G=0$, 
there is a one-to-one correspondence between
non-zero multipliers $Q|_\Esp$ and non-trivial conserved currents $(T,\Phi)|_\Esp$
modulo trivial ones.

All multipliers are given by the solutions of the determining equation
\begin{equation}\label{Qdeteqn}
0 = E_u(QG) = Q'{}^*(G) + G'{}^*(Q)
\end{equation}
where $E_u$ is the Euler-Lagrange operator (variational derivative), 
and where a prime denotes the Frechet derivative (linearization operator)
and a star denotes its adjoint.
For each solution $Q$,
a conserved current $(T,\Phi)|_\Esp$ can be obtained by several explicit methods \cite{Anc-review}.
Typically, the conservation laws having major physical interest,
such as mass, momentum, energy,
arise from multipliers of low order in derivatives of $u$ \cite{AncKar,Anc-review}. 

Now consider a PDE $G=0$ admitting a point symmetry
$\X=\tau(t,x,u)\partial_t + \xi(t,x,u)\partial_{x} +\eta(t,x,u)\partial_u$.
Any point symmetry yields a mapping $(t,x,u)\rightarrow (\tilde t,\tilde x,\tilde u)$
under which the space of conservation laws is mapped into itself,
preserving non-triviality.
In explicit form, the mapping 
$D_t T +D_x \Phi =0 \rightarrow D_{\tilde t}\tilde T +D_{\tilde x}\tilde \Phi=0$
is given by \cite{AncKar}
\begin{equation}\label{Xconslaw}
\begin{aligned}
\tilde T|_\Esp & =  \pr\X(T) + (D_x\xi + D_t\tau) T - (T D_t \tau + \Phi D_x\tau)
\\
\tilde \Phi|_\Esp & =  \pr\X(\Phi) + (D_x\xi + D_t\tau) \Phi - (T D_t \xi + \Phi D_x\xi)
\end{aligned}
\end{equation}
modulo a trivial current. 
Under this mapping,
the multiplier $Q\rightarrow \tilde Q$ is given by \cite{Anc2016,AncKar}
\begin{equation}\label{mappingQ}
\tilde Q = \pr\X(Q) + (R_\X +D_x\xi + D_t\tau) Q
\end{equation}
where $R_\X$ is the function defined by $\pr\X(G) =R_\X G$.

A conserved current is symmetry-invariant if
$(\tilde T,\tilde \Phi)|_\Esp$ vanishes. 
More generally, a conservation law is symmetry-invariant
if $(\tilde T,\tilde \Phi)|_\Esp$ is a trivial current. 
Symmetry-invariance of a conservation law is equivalent to the multiplier condition \cite{Anc2016,AncKar}
\begin{equation}\label{Qinvconslaw}
\tilde Q|_\Esp =0 . 
\end{equation} 

All conservation laws that are invariant under a given point symmetry $\X$
arise from multipliers that satisfy the determining equation \eqref{Qdeteqn} and the invariance condition \eqref{Qinvconslaw}.
These equations constitute a linear overdetermined system
whose solutions can be obtained by the same algorithm used to solve the well-known determining system for symmetries \cite{Olv-book,BCA-book}
\begin{equation}
\pr\X(G)|_\Esp =0 . 
\end{equation}

The set of all symmetry-invariant conservation laws under $\X$ is a subspace
in the space of all conservation laws admitted by a PDE $G=0$. 
If this invariant subspace has dimension $m\geq 1$,
then our reduction method will yield a set of $m$ first integrals
(some of which could be functionally dependent or trivially constant)
for the ODE given by symmetry-reduction of $G=0$.

\subsection{Travelling wave reduction}
\label{sec:travelwave-reduction}

A travelling wave solution of a PDE
$G(t,x,u,u_t,u_x,\ldots)$ $=0$
is of the form
\begin{equation}\label{travelwave:soln}
u(t,x)=U(x-ct)
\end{equation}
where $c=\const$ is the wave speed.
These solutions arise from invariance under the translation symmetry
\begin{equation}\label{travelwave:symm}
\X =\partial_t + c\partial_x
\end{equation}  
with $x-ct=\z$ and $u=U$ being the invariants of this symmetry. 
In a typical application, travelling wave solutions are sought for arbitrary $c\neq 0$.

Symmetry-invariance of the PDE is the condition that
$\pr\X(G)|_\Esp = (G_t + c G_x)|_\Esp =0$.
When the PDE is in a solved form for a leading derivative,
then the condition of symmetry-invariance can be usefully expressed in a simple way off of the solution space $\Esp$ of the PDE:
\begin{equation}\label{travelwave:symminvG}
\pr\X(G) = G_t + c G_x = 0 . 
\end{equation}    

Invariant solutions $u=U(\z)$ correspond to the reduction
$u(t,x) \rightarrow U(\z)$. 
These solutions satisfy the ODE obtained from reducing the PDE 
\begin{equation}\label{travelwave:ode}
G\big|_{u_t + c u_x=0}
=\barG(\z,\tfrac{dU}{d\z},\ldots) =  0
\end{equation}
which follows from symmetry-invariance \eqref{travelwave:symminvG}.
Here $u_t + c u_x=0$ is the invariant surface condition,
stating that the action of $\X$ on the function $u(t,x)$ vanishes. 

First integrals of the travelling wave ODE \eqref{travelwave:ode}
can be found by symmetry reduction of
all conservation laws that are invariant under the travelling-wave symmetry $\X$ of the PDE. 
The reduction can be understood through considering the action of $\X$ 
on a conservation law
expressed in terms of the canonical variables $(\z,\rho,U)$
for which $\X=\partial_\rho$,
where $\z=x-ct,U=u$ are the invariants and $\rho=x/c$ is a canonical coordinate.
In these variables,
the form of a conservation law $D_t T + D_x \Phi =0$
is $D_\z \barT + D_\rho \barPhi=0$,
with
\begin{equation}\label{travelwave:currentcanonical}
\barT = c T -\Phi,
\quad
\barPhi = -\tfrac{1}{c}\Phi 
\end{equation}
which are, in general, functions of $\rho,\zeta,U$ and derivatives of $U$.
Under the action of $\X=\partial_\rho$, 
the mapping of the conserved current $(\barT,\barPhi)$ consists of 
$\widetilde{\barT} = \pr\X(\barT) = \partial_\rho \barT$
and $\widetilde{\bar\Phi} = \pr\X(\barPhi) =\partial_\rho \barPhi$, 
which is shown by the form of the general mapping \eqref{Xconslaw}
with $(t,x,u)$ replaced by the canonical variables $(\z,\rho,U)$ of $\X$,
and with $(T,\Phi)$ replaced by the conserved current $(\barT,\barPhi)$ in canonical form. 
(In particular,
note $\X=\bar\tau\partial_\z +\bar\xi\partial_\rho + \bar\eta\partial_U$,
with $\bar\tau =\X\z =0$, $\bar\xi=\X\rho=1$, $\bar\eta=\X v=0$).
A conservation law is invariant under the travelling wave symmetry
iff $(\barT,\barPhi)$ is mapped into a trivial conserved current:
\begin{equation}\label{travelwave:symminv-canonical-conslaw}
\widetilde{\barT} = \pr\X(\barT) = \partial_\rho \barT = D_\rho\bar\Theta ,
\qquad
\widetilde{\bar\Phi} = \pr\X(\barPhi) =\partial_\rho \barPhi =-D_\z\bar\Theta .
\end{equation}

An equivalent and more useful formulation of this invariance condition \eqref{travelwave:symminv-canonical-conslaw}, 
as given by equations \eqref{mappingQ} and \eqref{Qinvconslaw}, 
is that 
\begin{equation}\label{travelwave:Qinvconslaw}
\pr\X(Q) = Q_t + c Q_x =0
\end{equation}
holds for the multiplier $Q$ of the conserved current $(T,\Phi)$. 

For a symmetry-invariant conservation law,
under symmetry reduction,
its conserved current in canonical form \eqref{travelwave:currentcanonical}
will have $\barT$ and $\bar\Phi$ given by functions of at most 
$\rho,\z,U$ and $\z$-derivatives of $U$.
Symmetry-invariance \eqref{travelwave:symminv-canonical-conslaw} then implies that
$\barT = \barTheta +\Psi$,
as obtained by integration with respect to $\rho$, 
where $\Psi$ is a function only of the invariants $\z,U$ and differential invariants $\tfrac{dU}{d\z},\tfrac{dU}{d\z^2},\ldots$.
Hence
$D_\z \barT = D_\z\barTheta + D_\z \Psi$ and $D_\rho \barPhi = -D_\z\barTheta$, 
and consequently 
\begin{equation}\label{travelwave:FIfromconslaw}
D_\z \barT + D_\rho \barPhi = D_\z \Psi =0 . 
\end{equation}

The reduced conservation law \eqref{travelwave:FIfromconslaw}
is a first integral $\Psi=\const$ of the travelling wave ODE \eqref{travelwave:ode}.
Its explicit form is given by the formula 
\begin{equation}\label{travelwave:FI}
\Psi(\z,U,\tfrac{dU}{d\z},\ldots)
= \barT + \smallint \partial_\rho \barPhi\; d\z
\end{equation}
which involves only the current
\begin{equation}\label{travelwave:transformedcurrent}
(\barT,\barPhi) =
(cT - \Phi,-\tfrac{1}{c}\Phi)\big|_{u(t,x)=U(\z),t=\rho -\z/c,x=c\rho} . 
\end{equation}
(The constant of integration with respect to $\z$ is fixed by requiring that $\rho$ does not appear in the righthand side of this formula \eqref{travelwave:FI},
namely $\partial_\rho(\barT + \smallint \partial_\rho\barPhi\; d\z)=0$.)

An equivalent formula for the first integral consists of 
\begin{equation}\label{travelwave:FIalt}
\Psi(\z,U,\tfrac{dU}{d\z},\ldots)
= \smallint \barQ\barG\; d\z
\end{equation}
in terms of a multiplier
\begin{equation}\label{travelwave:Qreduction}
\barQ = |J| Q\big|_{u(t,x)=U(\z),t=\rho-\z/c,x=c\rho}
\end{equation}
where $Q$ is the multiplier of the symmetry-invariant conservation law,
and where $|J|= \big|\frac{\partial(\z,\rho)}{\partial(t,x)}\big| =\tfrac{-1}{c}$ is a Jacobian factor 
arising from the point transformation $(t,x,u)\rightarrow (\z,\rho,U)$. 
In particular,
this formula \eqref{travelwave:FIalt} comes from the relation
$\barG\barQ = \frac{d\Psi}{d\z}$
which can be derived from combining
the reduction relations \eqref{travelwave:FIfromconslaw} and \eqref{travelwave:ode}. 

In summary,
each symmetry-invariant conservation law of a PDE $G=0$ 
reduces to a first integral of the travelling wave ODE $\barG=0$,
and all of these conservation laws can be found by
solving the multiplier determining equations \eqref{Qdeteqn} and \eqref{travelwave:Qinvconslaw}
for $Q$.
The resulting first integrals are then obtained
either from formula \eqref{travelwave:FI}
in terms of the conserved currents $(T,\Phi)$ arising from $Q$, 
or from the formula \eqref{travelwave:FIalt}
directly in terms of the reduced multipliers \eqref{travelwave:Qreduction}.  

\begin{exmp}\label{ex:compacteqn}
\emph{Compacton equation}

This example will show how our method is able to yield a quadruple reduction from a single symmetry.
It will also show how conservation laws containing $t$ and $x$ explicitly
may nevertheless reduce under translations to first integrals. 

Compactons are solitary travelling waves that have compact spatial support 
and exhibit robustness of their shape and speed in interactions \cite{RosZil}. 
They were discovered \cite{RosHym} for the $K(p,q)$ family of PDEs 
\begin{equation}\label{cpteqn}
u_t + a(u^p)_x +b(u^{q})_{xxx}=0,
\quad
p,q >0
\end{equation}
known as the compacton equation,
which was first motivated by the study of pattern formation in modelling liquid drops \cite{LudDra}.
Various aspects of this class of equations have continued to attract interest
\cite{LiOlvRos,AmbWri,HerHerEulEulRey},
and its conservation laws are well known \cite{RosZil}. 

Under the travelling wave symmetry \eqref{travelwave:symm}, 
the compacton equation \eqref{cpteqn} reduces to a third-order nonlinear ODE 
\begin{equation}\label{cpt:ODE}
bq U^{q-1} U''' + 3 bq(q-1) U^{q-2}U'U'' + bq(q-1)(q-2) U^{q-3}{U'}^3 + (ap U^{p-1}-c)U' =0
\end{equation}
for travelling wave solutions $u=U(\z)$, $\z=x-ct$. 
We will apply our reduction method to
obtain two functionally-independent first integrals
which reduce this ODE to first order.
In a special case of the compacton equation,
our reduction yields an additional first integral,
giving the explicit general solution of the ODE. 

Invariant conservation laws are determined by multipliers
$Q$ satisfying
\begin{equation}\label{cpt:invQeqn}
\pr\X(Q) = Q_t + c Q_x =0, 
\quad
E_u((u_t + a(u^p)_x +b(u^q)_{xxx})Q) =0 , 
\end{equation}
as shown by equations \eqref{Qdeteqn} and \eqref{travelwave:Qinvconslaw}.
Here we have used the fact that $R_\X =0$ for translation symmetries,
since the PDE \eqref{cpteqn} is in a solved form for $u_t$ which is a leading derivative. 

We will consider low-order multipliers: $Q(t,x,u,u_x,u_{xx})$. 
The determining equations \eqref{cpt:invQeqn}
split into an overdetermined linear system
which is straightforward to solve for $Q$ and $p,q,a,b$, with $p,q>0$ and $a,b\neq0$.
We obtain 
\begin{equation}\label{cpt:Qs}
Q_1=1,
\quad
Q_2= u^q , 
\end{equation}
holding for arbitrary $p,q,a,b$;
\begin{equation}\label{cpt:specialcaseQs}
Q_3= x-ct,
\quad
Q_4= (x-ct)^2 , 
\end{equation}
holding only when $p=1$, $a=c$. 

The resulting translation-invariant conservation laws are given by
\begin{align}
&\begin{aligned}
&
T_1=u , 
\\
& 
\Phi_1=a u^p + bq(q-1) u^{q-2}u_x^2 +bqu^{q-1}u_{xx} ; 
\end{aligned}
\label{cpt:invconslaw1}    
\\
&\begin{aligned}
&
T_2=\tfrac{1}{q+1} u^{q+1} , 
\\
&
\Phi_2=\tfrac{ap}{q+p} u^{p+q} +\tfrac{1}{2} b(q-2)q u^{2q-2}u_x^2+ b q u^{2q-1} u_{xx} ; 
\end{aligned}
\label{cpt:invconslaw2}
\end{align}
and, for the case $p=1$ and $a=c$,
\begin{align}
&\begin{aligned}
&
T_3=(x-ct)u , 
\\
&
\Phi_3= c(x-ct)u -bq u^{q-1}u_x +bq(q-1)(x-ct) u^{q-2}u_x^2 + bq(x-ct) u^{q-1} u_{xx} ; 
\end{aligned}
\label{cpt:invconslaw3}  
\\
&\begin{aligned}
&
T_4=(x-ct)^2 u , 
\\
&
\Phi_4= c (x-c t)^2 u +2 b u^q 
-2b q (x -ct) u^{q-1} u_x + b q(q-1)(x-c t)^2 u^{q-2} u_x^2 + b (x-c t)^2  u^{q-1} u_{xx} . 
\end{aligned}
\label{cpt:invconslaw4}  
\end{align}

From equation \eqref{travelwave:FI},
using the first two conservation laws \eqref{cpt:invconslaw1} and \eqref{cpt:invconslaw2},   
we get two first integrals
(which have been scaled by a constant factor for convenience)
\begin{align}
&
\Psi_1  = 
-c U + a U^p + b q(q-1) U^{q-2} {U'}^2 +b q  U^{q-1} U'' 
= C_1 , 
\label{cpt:FI1}
\\
&
\Psi_2  = 
(-\tfrac{c}{q+1} U +\tfrac{a p}{q+p} U^p)U^q
+\tfrac{1}{2} b q(q-2) U^{2 q-2} {U'}^2  + b q U^{2 q-1} U''
= C_2 . 
\label{cpt:FI2}
\end{align}
These first integrals are functionally independent.
Hence, when they are combined, 
this yields a reduced first-order ODE 
\begin{equation}
{U'}^2 =
\tfrac{2}{bq}\big(
\tfrac{c}{(q+1)}U^{3-q} -\tfrac{a}{q+p} U^{2+p-q} 
+ \tfrac{1}{q} U^{2-2q}(C_2 -C_1 U^q)
\big)
\end{equation}
which turns out to be separable. 
Hence,
we have obtained a triple reduction of the PDE \eqref{cpteqn} under a single symmetry,
leaving just a final quadrature to obtain the general travelling wave solution
for arbitrary $p,q,a,b$.
This quadrature contains the compacton as a special case. 

Furthermore,
the remaining two conservation laws \eqref{cpt:invconslaw3} and \eqref{cpt:invconslaw4}
which contain $t$ and $x$ explicitly in the combination $\z=x-ct$
also reduce to two first integrals through equation \eqref{travelwave:FI}. 
This yields 
\begin{align}
&\begin{aligned}  
\Psi_3  = &
b q U^{q-1} U' -b q(q-1) \z U^{q-2} U'^2 - b q\z U^{q-1} U''
= C_3 , 
\end{aligned}
\label{cpt:FI3}  
\\
&\begin{aligned}    
\Psi_4 = &
2b U^q - 2bq z U^{q-1} U'
+ bq(q-1) \z^2 U^{q-2} {U'}^2 + bq \z^2 U^{q-1} U''
= C_4
\end{aligned}
\label{cpt:FI4}
\end{align}
(which also have been scaled by a constant factor). 
Since we can specialize the previous first integrals \eqref{cpt:FI1} and \eqref{cpt:FI2} to this case $p=1$, $a=c$,
we have altogether four first integrals.
They satisfy the relation $2b\Psi_2 -\Psi_1 \Psi_4 -\Psi_3^2 =0$,
and so any three of them are functionally independent.
Combining $\Psi_1,\Psi_3,\Psi_4$, which are the three that have the simplest expressions,
we obtain
\begin{equation}\label{cpt:Usoln}
U(\z) =  b^{-\frac{1}{q}} (\tfrac{1}{2}C_1 \z^2 -C_3 \z +\tfrac{1}{2}C_4)^{\frac{1}{q}}
= (\tilde C_1 (\z +\tilde C_2)^2 +\tilde C_3)^{\frac{1}{q}}
\end{equation}
where $\tilde C_i$ ($i=1,2,3$) are arbitrary constants. 
Thus, with a single symmetry,
we have reduced the PDE \eqref{cpteqn} all the way to explicit quadrature
and obtained the general travelling wave solution \eqref{cpt:Usoln}
when $p=1$, $a=c$.
The PDE \eqref{cpteqn} in this case has the form
$u_t + c u_x +b(u^{q})_{xxx}=0$,
which possesses a compacton obtained by extending \eqref{cpt:Usoln} to $0$ for $|\z|>L$ 
with $\tilde C_2=0$, $\tilde C_1 =C>0$, $\tilde C_3=-CL^2$, and $q<\tfrac{1}{3}$.

\end{exmp}

\subsection{Similarity (scaling) reduction}
\label{sec:simil-reduction}

A similarity solution of a PDE
$G(t,x,u,u_t,u_x,\ldots)$ $=0$
is of the form
\begin{equation}\label{simil:soln}
u(t,x)= t^\alpha U(x/t^\beta) \text{ or }  u(t,x) = x^\alpha U(t/x^\beta) 
\end{equation}
where $\alpha,\beta=\const$ are scaling weights. 
These solutions arise from invariance under a scaling symmetry
\begin{equation}\label{simil:symm}
\X = t\partial_t + \beta x\partial_x +\alpha u\partial_u \text{ or }
\X = \alpha t\partial_t + x\partial_x +\beta u\partial_u 
\end{equation}  
with $x/t^\beta=\z$ and $u/t^\alpha =U$,
or $t/x^\beta=\z$ and $u/x^\alpha =U$, 
being the invariants of the symmetry. 
The two different forms are equivalent iff $\alpha,\beta\neq0$.
For simplicity, only the first form will be considered hereafter. 

Scaling-invariance of the PDE is the condition that
$\pr\X(G)|_\Esp = (t G_t + \beta x G_x + \alpha u G_u + (\alpha -1) u_t G_{u_t} + (\alpha -\beta) u_x G_{u_x} +\cdots)|_\Esp =0$.
When the PDE is in a solved form for a leading derivative,
then this symmetry invariance can be usefully expressed in a simple way off of the solution space $\Esp$ of the PDE:
\begin{equation}\label{simil:symminvG}
\pr\X(G) = t G_t + \beta x G_x + \alpha u G_u + (\alpha -1) u_t G_{u_t} + (\alpha -\beta) u_x G_{u_x} +\cdots = \Omega G
\end{equation}
where $\Omega=\const$ is the scaling weight of the PDE. 

Invariant solutions $u=t^\alpha U(\z)$, $\z=x/t^\beta$,
correspond to the reduction
$u(t,x) \rightarrow t^\alpha U(\z)$. 
These solutions satisfy the ODE obtained from reducing the PDE 
\begin{equation}\label{simil:ode}
t^{-\Omega} G\big|_{tu_t +\beta xu_x -\alpha u=0}
= \barG(\z,\tfrac{dU}{d\z},\ldots) =  0
\end{equation}
which follows from symmetry-invariance \eqref{simil:symminvG}.
Here $tu_t +\beta xu_x -\alpha u=0$ is the invariant surface condition,
stating that the action of $\X$ on the function $u(t,x)$ vanishes. 

First integrals of the similarity ODE \eqref{simil:ode} are inherited from
symmetry reduction of all conservation laws that are invariant under the scaling $\X$ of the PDE. 
Just like in the case of translation symmetries,
the reduction can be understood through
first introducing the canonical variables $(\z,\rho,U)$
for which $\X=\partial_\rho$,
where $\z=x/t^\beta,U=u/t^\alpha$ are the invariants and $\rho=\ln(t)$ is a canonical coordinate.
In these variables,
a conservation law $D_t T + D_x \Phi =0$ takes the form $D_\z \barT + D_\rho \barPhi=0$
given by 
\begin{equation}\label{simil:currentcanonical}
\barT = \beta x T - t\Phi,
\quad
\barPhi = -t^\beta T
\end{equation}
which are, in general, functions of $\rho,\zeta,U$ and derivatives of $U$.
The condition of symmetry-invariance of the transformed conservation law is given by the same formula \eqref{travelwave:symminv-canonical-conslaw}
as in the translation case.

The equivalent condition for symmetry-invariance
expressed in terms of the multiplier $Q$ of the conserved current $(T,\Phi)$
consists of 
\begin{equation}\label{simil:Qinvconslaw}
\pr\X(Q)  = t Q_t + \beta x Q_x + \alpha u Q_u + (\alpha -1) u_t Q_{u_t} + (\alpha -\beta) u_x Q_{u_x} +\cdots = -(\Omega + \beta + 1)Q . 
\end{equation}

Symmetry reduction of a scaling-invariant conservation law
yields a first integral $\Psi=\const$ of the similarity ODE, 
in the same way \eqref{travelwave:FIfromconslaw}
as for the case of translations. 
The explicit formula for the first integral is given by equation \eqref{travelwave:FI}
in terms of the current
\begin{equation}\label{simil:transformedcurrent}
(\barT,\barPhi) =
(\beta x T - t\Phi,-t^\beta T)\big|_{u(t,x)=t^\alpha U(\z),t=e^\rho,x=e^{\beta\rho}} . 
\end{equation}
An equivalent multiplier formula for the first integral comes from
combining 
the reduction relations \eqref{travelwave:FIfromconslaw} and \eqref{travelwave:ode}.
This yields equation \eqref{travelwave:FIalt} 
in terms of the multiplier
\begin{equation}\label{simil:Qreduction}
\barQ = |J| e^{\Omega\rho} Q\big|_{u(t,x)=t^\alpha U(\z),t=e^\rho,x=e^{\beta\rho}\z}
\end{equation}
where $Q$ is the multiplier of the scaling-invariant conservation law,
and where $|J|= \big|\frac{\partial(\z,\rho)}{\partial(t,x)}\big| =-e^{(\beta+1)\rho}$ is a Jacobian factor 
arising from the point transformation $(t,x,u)\rightarrow (\z,\rho,U)$. 

In summary,
each scaling-invariant conservation law of a PDE $G=0$ 
reduces to a first integral of the similarity ODE $\barG=0$,
and all of these conservation laws can be found by
solving the multiplier determining equations \eqref{Qdeteqn} and \eqref{simil:Qinvconslaw}
for $Q$.
The resulting first integrals are then obtained
either from formula \eqref{travelwave:FI}
in terms of the conserved currents \eqref{simil:transformedcurrent} arising from $Q$, 
or from the formula \eqref{travelwave:FIalt} 
directly in terms of the reduced multipliers \eqref{simil:Qreduction}. 

\begin{exmp}\label{ex:dBeqn}
\emph{Damped Boussinesq-type equation}

This example will show how our method is able to yield a double reduction
directly from a PDE,
without the necessity of any conservation laws being known a priori.

The Boussinesq equation is well-known to arise from the shallow water wave equations in a certain asymptotic regime.
When the effect of viscosity is considered,
which is relevant for modelling realistic wave motion, 
a damped Boussinesq equation is obtained \cite{Var1996}.
The study of the damped equation has attracted a lot of attention
(see, e.g. \cite{Var1997,Var1998,PolKayTut,LiFu}),
including double reduction under translations \cite{GanRos}.
There is particular interest in similarity solutions \cite{YanXieZha},
because in general they are connected to integrability features and blow-up solutions of nonlinear PDEs \cite{Cla,AblCla-book}. 

Here we look at the damped Boussinesq equation
with a $p$-power nonlinearity \cite{LiFu}
\begin{equation}\label{dBeqn}
u_{tt} - a u_{txx} + b u_{xxxx} = k(u^p)_{xx} ,
\quad
p\neq 0,1
\quad
a,b,k=\text{ positive }\const . 
\end{equation}
This equation possesses the scaling symmetry 
\begin{equation}\label{dB:scaling}
\X =2t\partial_t +x\partial_x- qu\partial_u,
\quad
q=\tfrac{2}{p-1} 
\end{equation}
which gives rise to similarity solutions
\begin{equation}\label{dB:similarity}
u(t,x) = t^{-\frac{q}{2}} U(\z), 
\quad
\z = x/\sqrt{t}
\end{equation}
where $\z$ and $U$ are the scaling invariants. 
The similarity ODE for these solutions is a fourth-order nonlinear ODE 
\begin{equation}\label{dB:ODE}
\begin{aligned}
0=
4b U'''' + 2a \z U''' +(\z^2 +\tfrac{4pa}{p-1} -4kp U^{p-1} )U''
&\\
-4kp(p-1) U^{p-2} {U'}^2   +\tfrac{3p+1}{p-1}\z U'+ \tfrac{4p}{(p-1)^2}U . 
\end{aligned}  
\end{equation}

We will apply our reduction method to
obtain a first integral which reduces this ODE to third order.
Unlike standard double reduction,
we do not need to know a scaling-invariant conservation law a priori,
and instead all such conservation laws will be derived
as a by-product of our method.

Scaling-invariant conservation laws are determined by multipliers
$Q$ satisfying
\begin{equation}\label{dB:invQeqn}
\pr\X(Q) + (q-1) Q = 0 ,
\quad
E_u((u_{tt} - a u_{txx} + b u_{xxxx} -k(u^p)_{xx})Q) =0 , 
\end{equation}
as shown by equations \eqref{Qdeteqn} and \eqref{Qinvconslaw}.
Here we have used 
\begin{equation}\label{dB:R}
\Omega =q-4
\end{equation}  
which is the scaling weight of the PDE \eqref{dBeqn}. 

We will consider multipliers of low order: 
$Q(t,x,u,u_t,u_x,u_{xx},u_{tx},u_{xxx})$. 
The determining equations \eqref{dB:invQeqn}
split into an overdetermined linear system
which is straightforward to solve for $Q$ and $p$, with $p\neq 0,1$.
This yields
\begin{align}\label{dB:Q}
& Q_1=1 ,
\quad
p=-1 ; 
\\
& Q_2=tx , 
\quad
p=2 ;
\\
& Q_3=t ,
\quad
p=3 . 
\end{align}

The resulting scaling-invariant conservation laws are given by:
\begin{align}
&\begin{aligned}
& T_1 = u_t , 
\\
& \Phi_1 = kb u_x/u^2 -au_{tx} + bu_{xxx} ; 
\end{aligned}
\label{dB:invconslaw1}  
\\
&\begin{aligned}
& T_2 = -xu +tx u_t , 
\\
& \Phi_2 = k t u^2 +a t u_t -2 k t x u u_x -a t x u_{tx} -b t u_{xx} +b t x u_{xxx} ; 
\end{aligned}
\label{dB:invconslaw2}  
\\
&\begin{aligned}
& T_3  = -u  + t u_t , 
\\
& \Phi_3 = -3 k t u^2 u_x -a t u_{tx}+ b t u_{xxx} . 
\end{aligned}
\label{dB:invconslaw3}
\end{align}
  
From equations \eqref{travelwave:FI} and \eqref{simil:transformedcurrent}, 
or alternatively from equation \eqref{simil:Qreduction}, 
we then get the first integrals
(which have been scaled by a constant factor for convenience)
\begin{align}
&\begin{aligned} 
\Psi_1 = & 
2 b U''' +a \z U'' +(\tfrac{1}{2}\z^2 +2k/U^2))U' -\tfrac{1}{2}\z U
= C_1,
\quad
p=-1 ; 
\end{aligned}
\\
&\begin{aligned}   
\Psi_2  = & 
2 b \z U''' +(a\z^2 -2b) U'' +(-4k \z U +\tfrac{1}{2}\z(\z^2+4a)) U'
+ 2(\z^2 -a) U  + 2 k U^2 = C_2,
\
p=2 ; 
\end{aligned}
\\
&\begin{aligned}   
\Psi_3  = & 
2 b U''' +a \z  U'' +(\tfrac{1}{2}\z^2 +2 a -6k U^2)U' +\tfrac{3}{2} \z U 
= C_3,
\quad
p=3 . 
\end{aligned}   
\end{align}
Thus, without having to know a conservation law a priori,
we have obtained a direct double reduction of damped Boussinesq equation \eqref{dBeqn} for similarity solutions with $p=-1,2,3$.

\end{exmp}

\section{Multi-reduction in $2+1$ dimensions under a solvable symmetry algebra}\label{sec:2+1}

The multi-reduction method described in the previous section
has a straightforward extension to reduction under a single symmetry of PDEs in more independent variables.
More importantly, 
it can be further generalized to reduction under solvable symmetry algebras. 
This goes beyond the scope of the standard double reduction method which is limited to single symmetries.
The most interesting situation is when the dimension of the symmetry algebra being used for reduction is one less than the number of independent variables in the PDE,
whereby the PDE will reduce to an ODE.
Each conservation law of the PDE that is invariant under the symmetry algebra
will reduce to a first integral of this ODE.
This leads to a reduction of the ODE by the number of first integrals that are functionally independent.

To illustrate this generalization,
we will explain how it works for two typical types of reductions for 
PDEs with a two-dimensional symmetry algebra in $2+1$ dimensions.
The first type of reduction will describe line travelling waves,
such as line solitons and eikonal waves, 
given by invariance under two commuting translations;
the second type will describe line similarity solutions
and similarity travelling waves, 
given by invariance under a scaling and a translation
which are not commuting but comprise a solvable algebra. 
These types of reductions are very common in applications,
and they will indicate how our method works for other abelian or solvable symmetry algebras
by going to suitably defined canonical variables. 

We begin by briefly discussing conservation laws, multipliers, and symmetries in $2+1$ dimensions. 

A local conservation law of a scalar PDE $G(t,x,y,u,u_t,u_x,u_y,\ldots)=0$ for $u(t,x,y)$ 
is a continuity equation $D_t T + D_x \Phi^x+ D_y \Phi^y =0$
holding on the space, $\Esp$, of solutions of the PDE,
where $T$ is the conserved density and $\Phi=(\Phi^x,\Phi^y)$ is the spatial flux vector,
which are functions of $t,x,y,u$, and derivatives of $u$.
The conserved current is $(T,\Phi)$.

A conserved current is trivial if it has the form
$T|_\Esp=D_x\Theta^x + D_y\Theta^y$
and $\Phi|_\Esp=(-D_t\Theta^x +D_y\Gamma,-D_t\Theta^y -D_x\Gamma)$, 
with $\Gamma$ and $\Theta=(\Theta^x,\Theta^y)$ being functions of $t,x,y,u$, and derivatives of $u$.

Just like in $1+1$ dimensions,
every non-trivial conservation law of the PDE $G=0$ arises from a multiplier, 
and there is a one-to-one correspondence between 
non-trivial conserved currents $(T,\Phi)|_\Esp$ modulo trivial ones
and non-zero multipliers $Q|_\Esp$,
with $QG =D_t T + D_x\Phi^x +D_y\Phi^y$ holding as an identity.  
Here $Q$ is a function of $t,x,y,u$, and derivatives of $u$,
such that $Q|_\Esp$ is non-singular. 
All multipliers are given by the solutions of the determining equation \eqref{Qdeteqn}. 
For each solution $Q$,
a conserved current $(T,\Phi)|_\Esp$ can be obtained by several explicit methods \cite{Anc-review}.

When the PDE $G=0$ admits a point symmetry
$\X=\tau(t,x,y,u)\partial_t +\xi^x(t,x,y,u)\partial_{x}+ \xi^y(t,x,y,u)\partial_{y} +\eta(t,x,y,u)\partial_u$,
the space of conservation laws is mapped into itself,
preserving non-triviality.
The corresponding mapping on multipliers 
$Q\rightarrow \tilde Q$ is given by \cite{Anc2016,AncKar}
\begin{equation}\label{2+1mappingQ}
\tilde Q = \pr\X(Q) + (R_\X +D_x\xi^x + D_y\xi^y + D_t\tau) Q
\end{equation}
where $R_\X$ is the function defined by $\pr\X(G) =R_\X G$.
A conservation law is symmetry-invariant if the conserved current is mapped into a trivial current: 
$(T,\Phi)|_\Esp \rightarrow (\tilde T,\tilde \Phi)|_\Esp=(D_{\tilde x}\tilde\Theta^{\tilde x}+D_{\tilde y}\tilde\Theta^{\tilde y},-D_{\tilde t}\tilde\Theta^{\tilde x} +D_{\tilde y}\tilde\Gamma,-D_{\tilde t}\tilde\Theta^{\tilde y} -D_{\tilde x}\tilde\Gamma)$. 
Symmetry-invariance of a conservation law is equivalent to the multiplier condition \eqref{Qinvconslaw}. 
We remark that previous work on double reduction in multi-dimensions \cite{BokDweZamKarMah}
considered only strict invariance of conserved currents such that $\tilde\Theta|_\Esp=0$,
entailing an unnecessary loss of generality.

The set of all symmetry-invariant conservation laws under $\{\X_1,\X_2\}$
is a subspace in the space of all conservation laws admitted by a PDE $G=0$. 
If this invariant subspace has dimension $m\geq 1$,
then our reduction method will yield a set of $m$ first integrals
(some of which could be functionally dependent or trivially constant)
for the ODE given by symmetry-reduction of $G=0$.

\subsection{Reduction by two translations: line travelling waves}
\label{sec:linewave-reduction}

A line travelling wave is the $2$-dimensional analog of a plane wave,
where the wave fronts along which the wave has a constant amplitude 
are lines orthogonal to the wave's direction of propagation. 
These waves have form
\begin{equation}\label{linewave}
u(t,x,y)  =U(\mu x+\nu y -t)
\end{equation}
where $\mu=\const$ and $\nu=\const$ determine
the speed $c=1/\sqrt{\mu^2 + \nu^2}$ of the wave 
and its direction $\theta=\arctan(\nu/\mu)$ in the $(x,y)$-plane. 
The wave fronts are the lines $\mu x+\nu y =\sigma=\const$,
with $-\infty<\sigma<\infty$.

Invariance of a PDE $G(t,x,y,u,u_t,u_x,u_y,\ldots)=0$
under the pair of commuting translation symmetries
\begin{equation}\label{linewave:symms}
\X_1 = (\mu^2 + \nu^2)\partial_t + \mu\partial_x + \nu\partial_y, 
\quad
\X_2 =\nu \partial_x -\mu\partial_y
\end{equation}
gives rise to line travelling wave solutions,
with $\z=\mu x+\nu y -t$ and $u=U$ being the joint invariants.
In particular,
invariance with respect to the spatial translation $\X_2$
imposes the line form of the travelling wave. 
Sometimes line travelling waves are simply called travelling waves,
but in general, travelling waves in dimensions higher than $1$
can depend on coordinates transverse to the direction of propagation of the wave. 
An example is the similarity travelling waves considered in the next subsection.

The condition of symmetry-invariance of the PDE can be formulated as
\begin{equation}\label{linewave:symminvG}
\pr\X_1(G) = (\mu^2 + \nu^2) G_t + \mu G_x + \nu G_y =0,
\quad
\pr\X_2(G) = \nu G_x -\mu G_y =0,
\end{equation}    
holding off of the solution space
whenever the PDE is in a solved form for a leading derivative. 

Line travelling wave solutions $u=U(\z)$ correspond to the reduction
$u(t,x,y) \rightarrow U(\z)$
and satisfy the ODE obtained from reducing the PDE 
\begin{equation}\label{linewave:ode}
G\big|_{\nu u_x -\mu u_y=0,\; (\mu^2+\nu^2) u_t + \mu u_x + \nu u_y =0}
=\barG(\z,\tfrac{dU}{d\z},\ldots) =  0 . 
\end{equation}
This is a consequence of conditions \eqref{linewave:symminvG},
where $\nu u_x -\mu u_y=0$ and $(\mu^2+\nu^2) u_t + \mu u_x + \nu u_y =0$
are the invariant surface conditions stating that the action of $\X_2$ and $\X_1$ on the function $u(t,x,y)$ vanishes.

First integrals of the line travelling wave ODE \eqref{linewave:ode}
are inherited by symmetry reduction of 
all conservation laws that are invariant under both symmetries $\X_1$ and $\X_2$ of the PDE. 
Just like the case of reduction under a travelling wave symmetry in $1+1$ dimensions,
here the condition of symmetry invariance of a conservation law
can be expressed straightforwardly by going to canonical variables
\begin{equation}\label{2+1pt}
(t,x,y,u) \rightarrow (\z,\rho,\chi,v)
\end{equation}  
given by $\rho= c^2 t$ and $\chi = c^2(\nu x - \mu y)$
for which the pair of symmetries jointly take the form
\begin{equation}\label{2+1commutingXs}
\X_1=\partial_{\rho},
\quad
\X_2=\partial_{\chi}, 
\end{equation}
with these variables satisfying 
$\X_1\rho=1$, $\X_2\chi=1$, along with $\X_1\chi=0$, $\X_2\rho=0$.
This transformation \eqref{2+1pt} sends a conservation law
$D_t T + D_x \Phi^x + D_y \Phi^y =0$
to an equivalent canonical form
$D_\z \barT + D_\rho \barPhi^\rho + D_\chi \barPhi^\chi =0$,
with 
\begin{equation}\label{linewave:currentcanonical}
\barT = \tfrac{1}{c^2}(\mu \Phi^x + \nu \Phi^y -T),
\quad
\barPhi^\rho = T ,
\quad
\barPhi^\chi = \nu \Phi^x - \mu \Phi^y
\end{equation}
which are, in general, functions of $\rho,\chi,\z,U$ and derivatives of $U$.
Then the condition for a conservation law to be invariant under the symmetries \eqref{2+1commutingXs} is given by 
\begin{subequations}\label{2+1symminv-canonical1-conslaw}
\begin{align}
& \pr\X_1(\barT) = \partial_\rho \barT
= D_\chi\barTheta_1^\chi+D_\rho\barTheta_1^\rho,
\label{2+1symminv-canonical1-T}
\\
& \pr\X_1(\barPhi) =(\partial_\rho \barPhi^\rho,\partial_\rho \barPhi^\chi)
=(-D_\z\barTheta_1^\rho + D_\chi \hat\Gamma_1,-D_\z\barTheta_2^\chi -D_\rho \hat\Gamma_1) , 
\label{2+1symminv-canonical1-Phi}
\end{align}
\end{subequations}
and
\begin{subequations}\label{2+1symminv-canonical2-conslaw}
\begin{align}
& \pr\X_2(\barT) = \partial_\chi \barT
= D_\chi\barTheta_2^\chi+D_\rho\barTheta_2^\rho,
\label{2+1symminv-canonical2-T}
\\
& \pr\X_2(\barPhi) =(\partial_\chi \barPhi^\rho,\partial_\chi \barPhi^\chi)
=(-D_\z\barTheta_2^\rho + D_\chi \hat\Gamma_2, -D_\z\barTheta_2^\chi -D_\rho \hat\Gamma_1) . 
\label{2+1symminv-canonical2-Phi}
\end{align}
\end{subequations}
This invariance condition \eqref{2+1symminv-canonical1-conslaw}--\eqref{2+1symminv-canonical2-conslaw}
can be expressed a more useful form in terms of the conservation law multiplier $Q$:  
\begin{equation}\label{linewave:Qinvconslaw}
\pr\X_1(Q) = (\mu^2 + \nu^2) Q_t + \mu Q_x + \nu Q_y =0,
\quad
\pr\X_2(Q) = \nu Q_x -\mu Q_y =0,
\end{equation}
which follows from equation \eqref{Qinvconslaw} applied to both symmetries
$\X=\X_1,\X_2$. 

Hence, all conservation laws that are invariant under the two symmetries
can be directly obtained from multipliers $Q$ that satisfy 
the invariance conditions \eqref{linewave:Qinvconslaw}
and the determining equation \eqref{Qdeteqn}. 
These three equations constitute a linear overdetermined system,
which can be solved by computer algebra analogously to the linear overdetermined system for symmetries.

In the same way as for reduction of conservation laws in $1+1$ dimensions
under a travelling wave symmetry,
every invariant conservation law 
$D_\z \barT + D_\rho \barPhi^\rho + D_\chi \barPhi^\chi =0$
reduces to $D_\z \Psi =0$ which is a first integral
$\Psi=\const$ of the line travelling wave ODE \eqref{linewave:ode}.
The explicit formula for the first integral is given by
\begin{equation}\label{2+1FI}
\Psi(\z,U,\tfrac{dU}{d\z},\ldots)
= \barT + \smallint (\partial_\rho \barPhi^\rho +\partial_\chi \barPhi^\chi)\; d\z
\end{equation}
in terms of the conserved current
\begin{equation}\label{linewave:transformedcurrent}
(\barT,\barPhi^\rho,\barPhi^\chi) =
(\tfrac{1}{c^2}(\mu \Phi^x + \nu \Phi^y -T),T,\nu \Phi^x - \mu \Phi^y)
\Big|{}_{\substack{ u(t,x,y)=U(\z)\hfill\\t=\frac{1}{c^2}\rho,x=c^2\mu\z +\mu\rho +\nu\chi,y=c^2\nu\z +\nu\rho -\mu\chi}} . 
\end{equation}
(The constant of integration with respect to $\z$ is fixed by requiring that both $\rho$ and $\chi$ do not appear in the righthand side of this formula \eqref{2+1FI}.)
Similarly to the situation in $1+1$ dimensions, 
there is an equivalent multiplier formula for the first integral,
which is given by 
\begin{equation}\label{2+1FIalt}
\Psi(\z,U,\tfrac{dU}{d\z},\ldots)
= \smallint \barQ\barG\; d\z , 
\end{equation}
using a reduced multiplier which is given by 
\begin{equation}\label{2+1Qreduction}
\barQ = (|J| Q)\Big|{}_{\substack{ u(t,x,y)=U(\z)\hfill\\t=\frac{1}{c^2}\rho,x=c^2\mu\z +\mu\rho +\nu\chi,y=c^2\nu\z +\nu\rho -\mu\chi}}
\end{equation}
in terms of the multiplier $Q$ of the symmetry-invariant conservation law
and a Jacobian $J= \frac{\partial(\z,\rho,\chi)}{\partial(t,x,y)}$ 
arising from the point transformation \eqref{2+1pt}.

\begin{exmp}\label{ex:gkp}
\emph{Generalized KP equation}

This example illustrates our reduction method for two commuting symmetries
and shows how it can lead to a triple reduction. 

The KP equation \cite{KadPet}
is an integrable generalization of the Korteweg-de Vries (KdV) equation in $2+1$ dimensions,
which arises in many important physical applications such as
shallow water waves \cite{KadPet,AblSeg1979}, 
matter-wave pulses in Bose-Einstein condensates \cite{HuaMakVel}, 
ion-acoustic waves in plasmas \cite{InfRow}, 
and ferromagnets \cite{Leb}. 

There has been recent interest in studying the KP equation with a $p$-power nonlinearity 
\begin{equation}\label{gkpeqn}
(u_t+ u^p u_x +u_{xxx})_x+ \sig u_{yy}=0,
\quad
\sig=\pm 1,
\quad
p>1 . 
\end{equation}
This gKP equation arises in modelling dispersive higher-order nonlinear wave phenomena
with small transverse modulations \cite{PelSteKiv,KarBel,InfRow}.
It possesses line solitons which are $2$-dimensional counterparts of the 
familiar solitons of the KdV equation.
Analysis of their kinematical features, stability, and other critical behaviour
has attracted considerable attention \cite{Sau,Liu,DubGra,AncGanRec2018}.

A line soliton is a solitary travelling wave of the line form \eqref{linewave},
with the wave amplitude $U(\z)$ satisfying asymptotic decay conditions
$U,U',U'',U'''\rightarrow 0$ as $|\z|\rightarrow \infty$.
Since it turns out that the line solitons of the gKP equation can never propagate
purely in the $y$ direction,
it will be useful to take $\z=x+\mu y-\nu t$, 
where the speed and direction of the soliton are given by
$c=\nu/\sqrt{\mu^2+1}$ and $\theta=\arctan(\mu)$, respectively.

These solutions arise from invariance under the pair of translation symmetries
\begin{equation}\label{gkp:symms}
\X_1 =\partial_t + \nu\partial_x , 
\quad
\X_2 =\mu \partial_x -\partial_y,
\end{equation}
whose joint invariants are $\z,U$. 
The line soliton ODE is given by
\begin{equation}\label{gkp:ODE}
U'''' +(\sig \mu^2 -\nu+U^p )U'' +pU^{p-1}U'{}^2 =0 . 
\end{equation}

Here, by working with the gKP equation in potential form
\begin{equation}\label{gkppoteqn}
v_{tx}+ (v_x)^p v_{xx} +v_{xxxx} + \sig v_{yy}=0,
\quad
u=v_x ,
\end{equation}
we will apply our reduction method supplemented by the shift symmetry
\begin{equation}\label{gkp:potsymm}
\X_3=\partial_v
\end{equation}  
which leaves invariant the relation between $u$ and the potential $v$,
and which commutes with the two translation symmetries \eqref{gkp:symms}.
This leads to two functionally independent first integrals of the fourth-order ODE \eqref{gkp:ODE},
where the first integrals are functions only of $V'=U,V''=U',V'''=U''$.
Hence, when the first integrals are combined,
we obtain a reduction of the ODE to a quadrature for $U(\z)$. 

Conservation laws that are invariant
under the symmetries \eqref{gkp:symms} and \eqref{gkp:potsymm} of the PDE \eqref{gkppoteqn}
are determined by multipliers $Q$ satisfying
\begin{equation}\label{gkp:invQeqn}
\begin{gathered}
\pr\X_1(Q) = Q_t + \nu Q_x =0, 
\quad
\pr\X_2(Q) = \mu Q_x - Q_y =0,
\quad
\pr\X_3(Q) = Q_v =0,
\\
E_v((v_{tx}+ (v_x)^p v_{xx} +v_{xxxx} + \sig v_{yy})Q)=0 . 
\end{gathered}
\end{equation}

We will consider low-order multipliers: $Q(t,x,v,v_t,v_x,v_y,v_{xx},v_{xxx})$. 
The determining equations \eqref{gkp:invQeqn}
split into an overdetermined linear system
which is straightforward to solve for $Q$ and $p$, with $p>0$ and $\sig=\pm1$.
We obtain four multipliers
\begin{equation}\label{gkp:Qs}
Q_1 =1,
\quad
Q_2=v_t,
\quad
Q_3=v_x,
\quad
Q_4=v_y , 
\end{equation}
holding for arbitrary $p$. 
The resulting symmetry-invariant conservation laws are given by
\begin{align}
&\begin{aligned}
T_1= v_x,
\quad
\Phi^x_1= v_{xxx}+\tfrac{1}{1+p}v_x^{1+p},
\quad
\Phi^y_1=v_y ; 
\end{aligned}
\label{gkp:invconslaw1}
\\
&\begin{aligned}
&
T_2 = \tfrac{1}{2}v_{xx}^2-\tfrac{1}{2}\sig v_y^2 -\tfrac{1}{(p+1)(p+2)}v_x^{p+2},
\\
&\qquad
\Phi^x_2 = v_t v_{xxx}-v_{tx}v_{xx}+\tfrac{1}{p+1} v_x^{p+1}v_t +\tfrac{1}{2}v_t^2,
\quad
\Phi^y_2= \sig v_t v_y ; 
\end{aligned}
\label{gkp:invconslaw2}
\\
&\begin{aligned}
&
T_3 = \tfrac{1}{2}v_x^2,
\quad
\Phi^x_3 = v_xv_{xxx}-\tfrac{1}{2}v_{xx}^2+\tfrac{1}{p+2} v_x^{p+2} -\tfrac{1}{2}\sig v_y^2,
\quad
\Phi^y_3 = \sig v_xv_y ; 
\end{aligned}
\label{gkp:invconslaw3}
\\
&\begin{aligned}
&
T_4 = \tfrac{1}{2}v_xv_y,
\quad
\Phi^x_4 = v_yv_{xxx}-v_{xx}v_{xy}+\tfrac{1}{p+1}v_y v_x^{p+1}+\tfrac{1}{2}v_t v_y,
\\
& \qquad
\Phi^y_4 = \tfrac{1}{2}v_{xx}^2+\tfrac{1}{2}\sig v_y^2 -\tfrac{1}{(p+1)(p+2)} v_x^{p+2} -\tfrac{1}{2}v_t v_x . 
\end{aligned}
\label{gkp:invconslaw4}
\end{align}
Note that the three conservation laws \eqref{gkp:invconslaw2}--\eqref{gkp:invconslaw4} 
are nonlocal in terms of $u$ but do not contain $v$ explicitly
(namely, just $v_t,v_y,v_x=u$ and their derivatives appear).

From equation \eqref{2+1FI},
we obtain four first integrals in terms of $V'=U,V''=U',V'''=U''$. 
However, only two of them are functionally independent, because
$\Psi_3=\mu \Psi_2$, $\Psi_1=-\nu \Psi_2$,
where
\begin{align}
\Psi_2 & = UU'' -\tfrac{1}{2}U'{}^2+\tfrac{1}{2}(\sig\mu^2-\nu)U^2 +\tfrac{1}{p+2}U^{p+2}
= C_2 , 
\label{gkp:FI1}
\\
\Psi_4 & = U'' +(\sig\mu^2-\nu)U +\tfrac{1}{p+1}U^{p+1}
= C_4 . 
\label{gkp:FI2}
\end{align}
By combining these first integrals,
and using the asymptotic decay conditions to find $C_2=C_4=0$, 
we get a reduced first-order ODE
\begin{equation}
{U'}^2 +(\sig\mu^2-\nu)U^2 +\tfrac{2}{(p+1)(p+2)}U^{p+2}= 0
\end{equation}
which is separable.
Thus, this yields a straightforward quadrature
giving the general solution of the line soliton ODE \eqref{gkp:ODE}.
See \Ref{AncGanRec2018} for more details,
and \Ref{AncRecGan2019a} for a related application. 

\end{exmp}

\subsection{Reductions by a scaling and a translation: line similarity solutions and travelling similarity solutions}
\label{sec:line-travel-similreductions}

A line similarity solution is a $2$-dimensional version of a similarity solution \eqref{simil:soln}
in which 
\begin{equation}\label{linesimil}
u(t,x,y)= t^\alpha U((\mu x+\nu y)/t^\beta) 
\end{equation}
has a constant amplitude along the spatial lines $\alpha x+\beta y =\sigma=\const$,
with $-\infty<\sigma<\infty$.
Here $\alpha,\beta=\const$ are scaling weights,
while $\mu,\nu=\const$ determine the direction of the lines. 
This type of solution arises when a PDE $G(t,x,y,u,u_t,u_x,u_y,\ldots)=0$
is invariant under a pair of symmetries
consisting of a spatial translation and a scaling
\begin{equation}\label{linesimil:symms}
\X_1 = \nu \partial_x -\mu \partial_y,
\quad
\X_2 = t\partial_t + \beta x\partial_x +\beta y\partial_y + \alpha u\partial_u , 
\end{equation}  
whose joint invariants are $(\mu x+\nu y)/t^\beta=\z$ and $u/t^\alpha =U$.

A travelling similarity solution, in contrast, is of the form
\begin{equation}\label{travelsimil}
u(t,x,y)= (\nu x -\mu y)^\alpha U((\mu x+\nu y -t)/(\nu x -\mu y)^\beta)
\end{equation}
which is a travelling wave with speed $c=1/\sqrt{\mu^2 + \nu^2}$
and direction $\theta=\arctan(\nu/\mu)$ determined by $\mu,\nu=\const$,
while $\alpha,\beta=\const$ are scaling weights.
This type of travelling wave differs from a line travelling wave \eqref{linewave}
by having a non-constant amplitude along the lines
$\mu x+\nu y=\sigma=\const$, $-\infty<\sigma<\infty$,
which are orthogonal to the wave's direction of propagation. 
In particular,
this amplitude has a similarity form in terms of $\nu x -\mu y$
which represents a coordinate on the lines. 
A PDE $G(t,x,y,u,u_t,u_x,u_y,\ldots)=0$ will possess travelling similarity solutions when it is invariant under the pair of symmetries 
\begin{equation}\label{travelsimil:symms}
\begin{aligned}
\X_1 = (\mu^2 + \nu^2)\partial_t + \mu\partial_x + \nu\partial_y, 
\quad
& 
\X_2 =  \beta t\partial_t + \tfrac{1}{\mu^2+\nu^2}\big( ((\beta\mu^2 +\nu^2)x +(\beta-1)\mu\nu y)\partial_x 
\\&\quad\qquad
+ ((\mu^2 +\beta\nu^2)y +(\beta-1)\mu\nu x)\partial_y \big) + \alpha u\partial_u , 
\end{aligned}
\end{equation}
representing a travelling-wave translation and a scaling aligned with the translation, 
whose joint invariants are
$(\mu x+\nu y -t)/(\nu x -\mu y)^\beta=\z$ and $u/(\nu x -\mu y)^\alpha =U$.

For both types of solutions,
the pair of symmetries comprises a solvable algebra
with the commutator structure
\begin{equation}\label{commstruct}
[\X_1,\X_2]=C X_1
\end{equation}
where $C=\beta=\const$ is the structure constant. 
To set up our reduction method,
it is useful to introduce two canonical variables $\rho,\chi$,
satisfying $\X_1\rho=1$, $\X_2\chi=1$.
These variables, however, will be not joint canonical coordinates
due to the lack of commutativity of the symmetries,
but we can impose the conditions 
$\X_1\chi=0$, $\X_2\rho= C\rho$,
whereby 
$\X_1=\partial_{\rho}$, $\X_2=\partial_{\chi} +C\rho\partial_{\rho}$
after use of a point transformation \eqref{2+1pt}.
This formulation will allow the remaining steps in our reduction method to carry through with little change compared to the previous case of commuting symmetries. 

One interesting aspect is the condition for symmetry-invariance \eqref{travelsimil:symms} of multipliers:
\begin{align}
& \begin{aligned}
& (\mu^2 + \nu^2) Q_t + \mu Q_x + \nu Q_y =0,
\end{aligned}
\label{2+1mappingQ-solvableX1}
\\
& \begin{aligned}
& \beta t Q_t + \tfrac{1}{\mu^2+\nu^2}\big( ((\beta\mu^2 +\nu^2)x +(\beta-1)\mu\nu y) Q_x
 + ((\mu^2 +\beta\nu^2)y +(\beta-1)\mu\nu x) Q_y \big)
\\&
+ \alpha u Q_u +\cdots  + (\Omega + 2\beta + 1)Q =  CQ 
  \end{aligned}
\label{2+1mappingQ-solvableX2}
\end{align}
where $\Omega=\const$ is the scaling weight of the PDE,
namely $\pr\X_2(G)=\Omega G$. 
The second invariance condition contains the extra term $CQ$,
which cancels the $\beta Q$ term. 
This extra term is related to the solvable structure of the symmetry algebra,
and it vanishes in the abelian case, $C=0$.

Nevertheless, the first integral formulas look the same as the ones
\eqref{2+1FI}, \eqref{2+1FIalt}, \eqref{2+1Qreduction}
derived in the abelian case. 

A final worthwhile remark about our method is that 
the symmetry reduction is accomplished in a single step, 
similarly to the abelian case,
without needing to do the reduction sequentially
(namely, it is not necessary to reduce the PDE and conservation laws
first under $\X_1$, show that $\X_2$ is inherited, and then reduce under $\X_2$.).

\begin{exmp}\label{ex:thinfilm}
\emph{Anisotropic thin film equation}

This example will show the situation where a PDE itself is in the form of a conservation law which will be used in the reduction,
and also how a classification problem is handled when a PDE contains a free function. 

The dynamics of a liquid film on a flat surface is modelled by \cite{FreSur}
the thin film equation, 
and its study is driven by many applications
in physics and chemistry \cite{OroDavBan,BecGru}
as well as in mathematics \cite{BerBow}. 
Here we look at an anisotropic version of the thin film equation 
\begin{equation}\label{thinfilmeqn}
u_t = -\tfrac{1}{k}( (u^3 (a_1 \Delta u -b_1 u^p)_x)_x + (u^3 (a_2\Delta u -b_2 u^p)_y)_y )
\end{equation}
$k=\const$ is the viscosity, 
$a_i,b_i=\const$ are coefficients for the anisotropic surface tension, 
$p$ is pressure nonlinearity power, 
and $\Delta = \partial_x^2 + \partial_y^2$ is Laplacian.
Note the isotropic case is given by $a_1=a_2$ and $b_1=b_2$. 

This equation \eqref{thinfilmeqn} has the form of a mass conservation law.
We consider line similarity solutions \eqref{linesimil},
which are relevant for understanding asymptotics and possible singular behaviour
in studying the equation.

We apply our reduction method using the pair of scaling and spatial translation symmetries \eqref{linesimil:symms}. 
It is easy to see that symmetry invariance of the PDE \eqref{thinfilmeqn} holds
iff $\alpha = \tfrac{-1}{2p+1}$ and $\beta = \tfrac{p-1}{2(2p+1)}$. 

The ODE for line similarity solutions is then found to be given by 
\begin{equation}
\big((\mu^2+\nu^2) (\mu^2 a_1 +\nu^2 a_2) U^3 U'''\big)'
-p\big((\mu^2 b_1 +\nu^2 b_2) U^{p+2} U'\big)' -\tfrac{k(p-1)}{2(2p+1)}\z U' 
-\tfrac{k}{2p+1} U =0 . 
\end{equation}

We will first take $Q=1$ for a multiplier,
corresponding to the mass conservation law.
The conditions \eqref{2+1mappingQ-solvableX1}--\eqref{2+1mappingQ-solvableX2}
for this conservation law to be invariant under the two symmetries \eqref{linesimil:symms} 
reduce to the algebraic condition $p=3$. 
In particular, this is the case for symmetry reduction of the PDE \eqref{thinfilmeqn} 
as a conservation law.

The resulting first integral is given by 
\begin{equation}
(\mu^2+\nu^2) (\mu^2 a_1 +\nu^2 a_2) U^3 U'''
-3(\mu^2 b_1 +\nu^2 b_2) U^{5} U' -\tfrac{k}{7}\z U = C_1 
\end{equation}
which provides a double reduction of the PDE in the case $p=3$. 

We can further apply our reduction method by searching for low-order multipliers
$Q(t,x,y,u,$ $u_x,u_y,u_{xx},u_{xy},u_{yy},$ $u_{xxx},u_{xxy},u_{xyy},u_{yyy})\neq\const$.
However, from the determining equations, 
we find that the thin film equation \eqref{thinfilmeqn}
possesses no other conservation laws that are invariant under the pair of
spatial translation and scaling symmetries \eqref{linesimil:symms}. 

\end{exmp}

\begin{exmp}\label{ex:semilinwave}
\emph{Wave equation with $p$-power nonlinearity}

This example will show how our reduction method can, in some situations,
yield sufficiently many first integrals to obtain the general solution
to the ODE for symmetry-invariant solutions in explicit form. 

The study of wave equations in $2+1$ dimensions is well-motivated
by obvious physical applications to vibrating membranes, flexural waves in thin plates, electromagnetic waves between two conducting plates, and so on.
Mathematical interest arises from the global behaviour of solutions,
especially the question of finite-time blow up or long-time existence
in the energy norm and in a Sobolev norm \cite{Sog}. 

Here we consider the wave equation with a power nonlinearity 
\begin{equation}\label{waveeqn}
u_{tt}- c^2(u_{xx} + u_{yy})  -ku^p =0,
\quad
p\neq 0,1
\end{equation}
where $c=\const$ is the wave speed
and $k\neq 0$ is the interaction constant.
Line travelling waves \eqref{linewave} of this equation
are the same as in 1+1 dimensions
(they are described by a nonlinear oscillator ODE and are well understood). 
More interesting are general travelling waves, which have the form
$u(t,x,y)=U(x-ct,y)$ after a rotation of the spatial coordinates $(x,y)$.
By rotational symmetry of equation \eqref{waveeqn}, 
there is no loss of generality in having a travelling wave move in any chosen direction.
We will specifically look at similarity travelling waves \eqref{travelsimil},
expressed as 
\begin{equation}\label{wave:travelsimil}
 u(t,x,y) = y^{\frac{2}{1-p}} U((x-\nu t)/y) 
\end{equation}
(namely, $\alpha=\tfrac{2}{1-p}$ and $\beta=1$). 

These solutions arise from invariance of the wave equation \eqref{waveeqn}
under a pair of symmetries given by a travelling wave translation and a scaling \eqref{travelsimil:symms} aligned along the $x$ direction:
\begin{equation}\label{wave:symms}
\X_1 = \partial_t + \nu\partial_x ,
\quad
\X_2 =  t\partial_t + x \partial_x + y\partial_y +  \tfrac{2}{p-1} u\partial_u . 
\end{equation}
The joint invariants of these symmetries are
$\z=(x-\nu t)/y$ and $U= u/y^\frac{2}{p-1}$, 
and the ODE for $U(\z)$ as obtained by reduction of the PDE \eqref{waveeqn}
is given by 
\begin{equation}\label{wave:ode}
(c^2(\z^2 +1)-\nu^2) U'' + 2c^2\tfrac{p+1}{p-1} \z U' + 2c^2 \tfrac{p+1}{(p-1)^2}U + kU^p =0 . 
\end{equation}

Conservation laws that are invariant under both symmetries of the PDE \eqref{waveeqn}
are determined by multipliers $Q$ satisfying
\begin{equation}\label{wave:invQeqn}
\begin{gathered}
\pr\X_1(Q) = Q_t + cQ_x =0,
\quad
\pr\X_2(Q) +(2\beta +1 -C) Q= \beta t Q_t + \beta x Q_x + y Q_y + \alpha u Q_u =0 , 
\\
E_u((u_{tt}- c^2(u_{xx} + u_{yy})  -ku^p)Q) =0 ,
\end{gathered}
\end{equation}
where $C=1$ is the structure constant in the symmetry algebra. 

We will consider low-order multipliers $Q(t,x,y,u,u_t,u_x,u_y)$.
The determining equations \eqref{wave:invQeqn}
split into an overdetermined linear system
which is straightforwardly solved to find $Q$ and $p$, with $p\neq 0,1$.
This yields two cases:
\begin{gather}
p=-3;
\quad
Q_1= u_t,
\quad
Q_2=u_x,
\quad
Q_3=u_y ; 
\\
p=5,
\quad
|\nu|=c;
\quad
Q_4=
(\nu t -x)^2(u_t -\nu u_x) + y^2( u_t +\nu u_x) +\nu(\nu t-x)(2 y u_y + u)  . 
\end{gather}
In fact, the first set of multipliers yield conservation laws for all $p$,
but invariance holds only for $p=-3$.
The resulting invariant conservation laws are given by
\begin{align}
&  T_1= \tfrac{1}{2}(u_t^2 +c^2(u_x^2 +u_y^2) -k/u^2),
\quad
\Phi_1^x = -c^2 u_t u_x ,
\quad
\Phi_1^y = -c^2 u_t u_y ;
\label{wave:travelsimil-conslaw1}
\\
& 
T_2= -\tfrac{1}{c^2} \Phi_1^x,
\quad
\Phi_2^x= T_1 - c^2 u_x^2 -u_t^2,
\quad
\Phi_2^y = -c^2 u_x u_y ; 
\label{wave:travelsimil-conslaw2}
\\
&
T_3= -\tfrac{1}{c^2} \Phi_2^x,
\quad
\Phi_3^x= -c^2 u_x u_y,
\quad
\Phi_3^y =T_1 -c^2 u_y^2 -u_t^2
\label{wave:travelsimil-conslaw3}
\end{align}
which represent energy and $x,y$-momentum;
and in addition 
\begin{equation}\label{wave:travelsimil-conslaw4}
\begin{aligned}
& T_4 =
((\nu t -x)^2+y^2) T_1
+\tfrac{1}{\nu} ((\nu t -x)^2 -y^2) \Phi_1^x 
+\tfrac{2}{\nu}y(x-\nu t)\Phi_1^y 
+\nu(x-\nu t)u u_t +\tfrac{1}{2}c^2 u^2,
\\
& \Phi_4^x =
\nu(y^2 -(x-\nu t)^2)\Phi_2^x 
+((x-\nu t)^2+y^2)\Phi_1^x
-2\nu y(x-\nu t)\Phi_3^x
-\nu^3(x-\nu t)u u_x +\tfrac{1}{2}\nu^3 u^2,
\\
& \Phi_4^y =
\nu(y^2 -(x-\nu t)^2)\Phi_2^y
+((x-\nu t)^2+y^2)\Phi_1^y
-2\nu y(x-\nu t)\Phi_3^y
-\nu^3(x-\nu t)u u_y , 
\end{aligned}
\end{equation}
which represents a conformal energy.
The relation between the various components here
is a consequence of Lorentz invariance of the PDE \eqref{waveeqn}. 

From the first three conservation laws \eqref{wave:travelsimil-conslaw1}--\eqref{wave:travelsimil-conslaw3}, 
we get three first integrals,
but only two of them are functionally independent:
\begin{align}
\Psi_1 & =
(\nu^2 -c^2(\z^2+1)) {U'}^2 +\tfrac{1}{4}c^2 U^2 +k U^2
= C_1 = -\tfrac{1}{\nu}C_3 ; 
\\
\Psi_2 & =
(\nu^2 - c^2(\z^2+1))\z (U'-\tfrac{1}{2} U/\z)^2+ \tfrac{1}{4}(c^2 -\nu^2)U^2/\z+ k\z/U^2 = C_2 . 
\end{align}
By combining and re-scaling these first integrals,
we obtain 
\begin{equation}\label{wave:gensoln}
U(\z)^2 =C_1 + C_2 \z \pm\tfrac{1}{c}\sqrt{ (c^2(\z^2 +1)- \nu^2)(C_2^2 +(4k -c^2C_1^2)/(\nu^2-c^2)) } . 
\end{equation}
This yields the general solution of the similarity travelling wave ODE \eqref{wave:ode} when $p=-3$.
Interestingly, this power turns out to be the critical power (for scaling invariance) of the Sobolev norm for $H^{\frac{3}{2}}$.
Further discussion of the solution \eqref{wave:gensoln} will be given elsewhere.

Finally, from the fourth conservation law \eqref{wave:travelsimil-conslaw4},
we find that its reduction vanishes,
namely, it yields a trivial first integral, $\Psi_4=0$. 
The power $p=5$ is recognized to be the critical power (for scaling invariance) of the energy norm
defined by the conserved energy integral $\int_\Rnum T_1\;dx\;dy$. 
  
\end{exmp}

\section{Multi-reduction under a non-solvable symmetry algebra}{}\label{sec:n+1}

The multi-reduction method outlined in the previous section
can be applied to any symmetry algebra that has a solvable structure. 
It can also be applied more generally to non-solvable symmetry algebras
under certain conditions.
For reduction of a PDE with $n$ independent variables to an ODE, 
the symmetry algebra needs to have two invariants,
one of which must involve the dependent variable in the PDE.
A more precise statement of the theory will be given elsewhere \cite{AncGan2019}.
Similarly to the case of a solvable symmetry algebra,
the ODE will inherit a first integral from each symmetry-invariant conservation law of the PDE. 
This provides a further reduction of the ODE
by the number of first integrals that are functionally independent.

A typical example is reduction under of a PDE in $n+1$ dimensions
under the group of spatial rotations $SO(n)$ combined with an additional symmetry such as a scaling.

\begin{exmp}\label{ex:porousmedia}
\emph{Porous medium  equation}

Fluid flow in a porous medium,
and heat transfer in conducting media such as plasmas, 
as well as other nonlinear diffusion processes 
can be modelled by \cite{Bar1987,Vaz}
\begin{equation}\label{porousmedeqn}
u_t = k \Delta (u^p)
\end{equation}
known as the porous medium equation, 
where $k>0$ is the diffusion constant, $p\neq 0,1 $ is a nonlinearity power
and $\Delta$ is the Laplacian.
It is of interest to study this equation in $n$ spatial dimensions,
for $u(t,\x)$ with $\x\in\Rnum^n$,
and there has been considerable mathematical attention to question of global behaviour of solutions \cite{Vaz}.

In particular, exact solutions have been found \cite{Bar1987,Puk},
one of which represents the fundamental solution describing the dynamics of
a point-like source at $t=0$. 
This solution satisfies an ODE obtained by reduction of the porous medium equation
under the $SO(n)$ group of rotations together with a certain scaling group. 

The rotations and scalings are symmetries whose generators
\begin{equation}\label{porousmed:symms}
\begin{aligned}
& \X_1 = \boldsymbol{\xi}\cdot\partial_\x,
\quad
\boldsymbol{\xi} = \A\cdot\x,
\quad
\A^\t=-\A , 
\\
& X_2= a t\partial_t + \x\cdot\partial_\x + q u\partial_u,
\quad
a= q(1-p) +2 , 
\end{aligned}
\end{equation}  
comprise a one-parameter family of non-solvable algebras
$\mathfrak{so}(n)\times \Rnum$.
Here $q$ is a scaling weight
which parameterizes the family of scalings;
$\A$ is an arbitrary constant antisymmetric matrix
which represents the $2$-dimensional plane in which a rotation takes place.
Any choice of basis for the vector space of all matrices $\{\A\}$
corresponds to a basis for $\mathfrak{so}(n)$.

The joint invariants of the symmetry algebra (for any fixed $q$)
are $\z=r/t^{\frac{1}{a}}$ and $U=u/t^{\frac{q}{a}}$,
with $r=|\x|$ being the radius variable. 
Symmetry reduction of equation \eqref{porousmedeqn} 
yields 
\begin{equation}
p (U^{p-1} U')' +(\tfrac{1}{m}z+ p(n-1)U^{p-1}/z)U' +\tfrac{n}{m}U =0, 
\quad
m = n(p-1)+2 
\end{equation}
which is a nonlinear second-order ODE for $U(\z)$.
The resulting symmetry-invariant solutions of equation \eqref{porousmedeqn}
have a rotationally-invariant similarity form
\begin{equation}\label{porousmed:invsol}
u(t,\x) = t^{\frac{q}{a}} U(r/t^{\frac{1}{a}}) . 
\end{equation}

It is well known \cite{Bar1987} that the space of non-trivial conservation laws
$D_t T + D_\x\cdot\Phi^\x=0$ of equation \eqref{porousmedeqn}
is isomorphic to solutions of $\Delta f(\x)=0$,
with their multipliers being given by $Q=f$.
(In particular, the determining equation
$E_u((u_t -k \Delta (u^p))Q)=0$ for $Q(t,\x)$ splits into $Q_t=0$, $\Delta Q=0$.)
The conservation laws that are invariant under all of the symmetries \eqref{porousmed:symms} (for fixed $q$)
are determined by 
\begin{equation}
\pr\X_1 Q = \boldsymbol{\xi}\cdot\nabla f =0,
\quad
\pr\X_2 Q = \x\cdot\nabla f + (a+n +\Omega)f = 0
\end{equation}
where $\Omega = pq-2$ is scaling weight of equation \eqref{porousmedeqn}. 

These determining equations split into an overdetermined linear system
consisting of $\x\wedge\nabla f=0$ and $\x\cdot\nabla f =-(q+n)f$,
with $\Delta f=0$.
All solutions are easily found to be given by 
\begin{align}
& q=-n:
\quad
Q_1= 1 ; 
\\
& q=-2,
\quad
n\neq 2 :
\quad
Q_2= r^{2-n} . 
\end{align}
They yield the symmetry-invariant conservation laws 
\begin{align}
& T_1= u,
\quad
\Phi_1^\x = -kp u^{p-1}\nabla u ; 
\\
& T_2 = r^{2-n} u,
\quad
\Phi_2^\x = -k r^{2-n} u^{p-1}(p r\nabla u + (n-2) u \x/r) . 
\end{align}
For solutions \eqref{porousmed:invsol},
we obtain first integrals
\begin{align}
& \Psi_1 = pm \z^{n-1} U^{p-1} U'+ \z^n U = C_1,
  \quad
  q = -n;
\label{porousmed:FI1}
\\
& \Psi_2 = p \z U^{p-1} U' +\tfrac{1}{2p} \z^2 U +(n-2) U^p = C_2,
  \quad
  q=-2,
  \quad
  n\neq 2 . 
\label{porousmed:FI2}
\end{align}
Thus, we have reduced the second-order PDE in $n+1$ dimensions
to a first-order ODE for rotationally-invariant similarity solutions 
$u(t,\x)= t^{-\frac{n}{m}}U(r/t^{\frac{1}{m}})$ and $u(t,\x)= t^{-2p}U(r/t^{2p})$,
respectively.
Note these two reductions are distinct for $p\neq1$. 

We also note that the known fundamental solution \cite{Bar1987}
in the case $1-\tfrac{2}{n}<p<1$ and $n\geq 2$
can be found by integration of the first integral \eqref{porousmed:FI1} with $C_1=0$,
yielding 
$U(\z) = U_0/(\z^2 +C_0)^{\frac{1}{1-p}}$, $U_0=\big(\tfrac{1-p}{2p(2+(p-1)n)}\big)^{\frac{1}{p-1}}$
where $C_0$ determines the total mass of the solution. 

\end{exmp}

\section{Conclusions}\label{sec:remarks}

First, starting from a point symmetry of a given PDE in two independent variables,
our basic result provides an explicit algorithmic method to find 
all symmetry-invariant conservation laws that will reduce to first integrals for
the ODE describing symmetry-invariant solutions of the PDE. 
This significantly generalizes the double reduction method known in the literature. 

Second, we have extended this multi-reduction method to PDEs that have $n>2$ independent variables and a symmetry algebra of dimension $n-1$. 
Our method then yields a direct reduction --- in one step --- of the PDE
to an ODE for invariant solutions plus a set of first integrals.
The symmetry algebra is not restricted to be solvable.
Moreover,
we formulate the condition of symmetry-invariance of a conservation law
in an improved way by using multipliers, 
thereby allowing symmetry-invariant conservation laws to be obtained algorithmically,
without the need to first find conservation laws and then check their invariance.
This cuts down considerably the number and complexity of computational steps involved in the reduction method. 

Third, if the space of symmetry-invariant conservation laws has dimension $m\geq 1$,
then the method yields $m$ first integrals
along with a check of which ones are non-trivial via their multipliers.
We have shown examples for which this reduction gives
the explicit general solution for symmetry-invariant solutions of the given PDE.

Our method also can be used to find all conservation laws inherited by
a PDE in fewer variables obtained from symmetry reduction of PDE in more than two independent variables.
This extension will be explained elsewhere \cite{AncGan2019}. 

All of these results are new and can be expected to have a wide applicability in the study of PDEs.

To illustrate the method,
we have considered several examples of interesting symmetry reductions:
travelling waves and similarity solutions in $1+1$ dimensions;
line travelling waves, line similarity solutions, and similarity travelling waves in $2+1$ dimensions;
rotationally symmetric similarity solutions in $n+1$ dimensions.

\section*{Acknowledgements}
SCA is supported by an NSERC Discovery Grant and thanks the University of C\'adiz for support during a visit when this work was completed. 
MLG thanks Cadiz University and FQM-201 Junta de Andalucia group for support.

\end{document}